\documentclass[leqno,12pt,twoside,a4paper]{article}
\usepackage{hyperref}
\usepackage{amsmath}
\usepackage[psamsfonts]{amssymb}
\usepackage{amsthm}
\usepackage[mathscr,psamsfonts]{eucal}
\usepackage{cmmib57}
\usepackage{amscd}
\input{ueur.fd}\input{ueur57.fd}
\DeclareMathAlphabet{\eurm}{U}{eur}{m}{n}
\DeclareMathAlphabet{\eubf}{U}{eur}{b}{n}
\DeclareFontFamily{OML}{cyr}{}
\DeclareFontShape{OML}{cyr}{m}{n}{
   <5> <6> <7> <8> <9>
   <10> <10.95> <12> <14.4> <17.28> <20.74> <24.88> wncyr10
  }{}
\DeclareSymbolFont{rusletters}{OML}{cyr}{m}{n}
\DeclareSymbolFontAlphabet{\rusmath}{rusletters}
\DeclareMathSymbol{\D}{\rusmath}{rusletters}{"64}
\newcounter{assump}
\newtheorem{Assumption}{\indent Assumption}[assump]
\newtheorem{Definition}{\indent Definition}[section]
\newtheorem{Lemma}{\indent Lemma}[section]
\newtheorem{Proposition}{\indent Proposition}[section]
\newtheorem{Theorem}{\indent Theorem}[section]
\newtheorem{Corollary}{\indent Corollary}[section]
\newtheorem{Remark}{\indent Remark}[section]
\newtheorem{Note}{\indent Note}[section]
\newtheorem{Example}{\indent Example}[section]
\newtheorem{Problem}{\indent Problem}[section]
\newcommand{\bAs}{\begin{Assumption}\em}
\newcommand{\eAs}{\end{Assumption}}
\newcommand{\bDf}{\begin{Definition}\em}
\newcommand{\eDf}{\end{Definition}}
\newcommand{\bLm}{\begin{Lemma}}
\newcommand{\eLm}{\end{Lemma}}
\newcommand{\bPr}{\begin{Proposition}}
\newcommand{\ePr}{\end{Proposition}}
\newcommand{\bTh}{\begin{Theorem}}
\newcommand{\eTh}{\end{Theorem}}
\newcommand{\bCr}{\begin{Corollary}}
\newcommand{\eCr}{\end{Corollary}}
\newcommand{\bRm}{\begin{Remark}\em}
\newcommand{\eRm}{\end{Remark}}
\newcommand{\bNt}{\begin{Note}\em}
\newcommand{\eNt}{\end{Note}}
\newcommand{\bEx}{\begin{Example}\em}
\newcommand{\eEx}{\end{Example}}
\newcommand{\bPb}{\begin{Problem}\em}
\newcommand{\ePb}{\end{Problem}}
\newcommand{\bPf}{\begin{proof}[\noindent\indent{\sc Proof}]}
\newcommand{\ePf}{\renewcommand{\qedsymbol}{}\end{proof}}
\newcommand{\bEq}{\begin{eqnarray}}
\newcommand{\eEq}{\end{eqnarray}}
\newcommand{\beq}{\begin{eqnarray*}}
\newcommand{\eeq}{\end{eqnarray*}}
\def\QED{\hskip0.1em\hfill\null\ \null\nobreak\hfill\kern3pt\vbox
{\hrule\hbox
   {\vrule\kern1pt\vbox{\kern1.7pt\hbox{$\scriptscriptstyle{QED}$}
    \kern0.2pt}\kern1pt\vrule}\hrule}}

\def\END{\hskip0.1em\hfill\null\ \null\nobreak\hfill\kern3pt\vbox
{\hrule\hbox
   {\vrule\kern1pt\vbox{\kern1.7pt\hbox{$\,\,\,\vspace{5pt}$}
    \kern0.2pt}\kern1pt\vrule}\hrule}}
\newcommand{\ie}{i.e$.$ }
\newcommand{\R}{I\!\!R}
\newcommand{\A}{\forall}

\newcommand{\mto}{\mapsto}
\newcommand{\hto}{\hookrightarrow}
\newcommand{\der}{\partial}
\newcommand{\nab}{\nabla}
\newcommand{\Nat}{^{\natural}}
\newcommand{\Fla}{^{\flat}}

\newcommand{\cin}{C^\infty}

\DeclareMathOperator{\byd}{\,{\raisebox{.1ex}{$\eurm :$}{\eurm =}}\,}

\newcommand{\car}{\times}

\newcommand{\sub}{\subset}

\newcommand{\wed}{\wedge}

\newcommand{\com}{\!\circ\!}
\newcommand{\con}{\,\lrcorner\,}
\newcommand{\ten}{\!\otimes\!}

\newcommand{\oten}[1]{\overset{#1}{\otimes}}
\newcommand{\uten}[1]{\underset{#1}{\otimes}}

\newcommand{\owed}[1]{\overset{#1}{\wedge}}

\newcommand{\oset}[2]{\overset{#1}{#2}}

\newcommand{\olin}[1]{\overline{#1}}

\newcommand{\alp}{\alpha}
\newcommand{\bet}{\beta}
\newcommand{\gam}{\gamma}

\newcommand{\eps}{\epsilon}

\newcommand{\lam}{\lambda}
\newcommand{\sig}{\sigma}

\newcommand{\ome}{\omega}
\newcommand{\Gam}{\Gamma}

\newcommand{\The}{\Theta}
\newcommand{\Lam}{\Lambda}

\newcommand{\Ome}{\Omega}

\newcommand{\varkap}{\varkappa}

\newcommand{\bB}{\boldsymbol{B}}

\newcommand{\bE}{\boldsymbol{E}}
\newcommand{\bF}{\boldsymbol{F}}
\newcommand{\bG}{\boldsymbol{G}}

\newcommand{\bM}{\boldsymbol{M}}

\newcommand{\bP}{\boldsymbol{P}}

\newcommand{\bR}{\boldsymbol{R}}
\newcommand{\bS}{\boldsymbol{S}}
\newcommand{\bT}{\boldsymbol{T}}

\newcommand{\cE}{\mathcal{E}}

\newcommand{\cH}{\mathcal{H}}

\newcommand{\cL}{\mathcal{L}}

\newcommand{\cP}{\mathcal{P}}

\newcommand{\BE}{{\mathbb{E}}}

\newcommand{\BL}{{\mathbb{L}}}
\newcommand{\BM}{{\mathbb{M}}}
\newcommand{\BN}{{\mathbb{N}}}

\newcommand{\BP}{{\mathbb{P}}}
\newcommand{\BQ}{{\mathbb{Q}}}
\newcommand{\BR}{{\mathbb{R}}}

\newcommand{\BT}{{\mathbb{T}}}

\newcommand{\myskip}{\vspace*{8pt}}

\newcommand{\uf}{\underline{f}}
\newcommand{\uX}{\underline{X}}

\newcommand{\h}{\hbar}

\newcommand{\of}{\oset{o}{f}}

\DeclareMathOperator{\Co}{\it Co}
\newcommand{\Fg}{\mathfrak{g}}

\title{\textbf{On symmetries in covariant Galilei 
               mechanics}}
\author{Dirk Saller
\\
{\small Department of Mathematics, Mannheim University}
\\
{\small D 7, 27 68131 Mannheim, Germany}
\\
{\small email: saller@euler.math.uni-mannheim.de}
\and
Raffaele Vitolo
\\
{\small Department of Mathematics `E. De Giorgi', University of Lecce}
\\
{\small via per Arnesano, 73100 Lecce, Italy}
\\
{\small email: Raffaele.Vitolo@unile.it}}
\date{}
\begin{document}

\maketitle

\begin{abstract}
In the framework of covariant classical mechanics (\ie, generally
relativistic classical mechanics on a spacetime with absolute time)
developed by Jadczyk and Modugno, we analyse systematically the
relations between symmetries of geometric objects.  We show that
the (holonomic) infinitesimal symmetries of the cosymplectic structure
and of its horizontal potentials are also symmetries of spacelike metric,
gravitational and electromagnetic fields, Euler-Lagrange morphism and
Lagrangians.  Then, we provide a definition for a covariant momentum map 
associated with a group of cosymplectic symmetries using a covariant lift 
of functions of phase space.  In the case when the cosymplectic 
symmetries projects on spacetime we see that the components of this 
momentum map are quantisable functions in the sense of Jadczyk and 
Modugno.  Finally, we illustrate the results by some examples.
\end{abstract}
\vspace*{3cm}
{\it Key words: Covariant classical mechanics, Covariant quantum 
mechanics,\\
\hspace*{2cm} Lie groups and symmetry,  Jet contact structure,
Cosymplectic manifolds}
\newpage
\tableofcontents
\section{Introduction}

At the beginning of the 90's M. Modugno and A. Jadczyk proposed a new 
geometric framework for a generally covariant classical and quantum 
mechanics on a curved spacetime with absolute time 
\cite{JadJanMod98,JadMod93,Mod99}, based on jets, 
connections and cosymplectic forms. 
This approach was later developed in
\cite{CJM95,Jan94,MoVi95,MTV99,Vit95,Vit96,Vit99}. 
The theory will be referred to as `covariant classical Galilei
theory' ({\em CCG}) and `covariant quantum Galilei 
theory' ({\em CQG}).  

The theory was partially inspired by the wide literature on
geometric  formulations of classical and quantum mechanics. Main
sources were symplectic and cosymplectic classical mechanics
\cite{Albert89,CdLL92,dLS93,Got86,LiMa87,Mar83,MaRa95,MasPag91},
Newton or  Galilei classical mechanics as general relativistic
theories 
\cite{LBLL73,Duv93,DBKP85,DuKu84,Ehl89,Kuc80,Kun84,Tra63,Tul85},
quantum theories of mechanics within a symplectic or cosymplectic
framework (such as geometric quantisation)
\cite{deLeon97,DuKu84,Got86,Kos70,LanLin91,Sni80,Sou70,Tul85,Woo92}.

The models of {\em CCG\/} and {\em CQG\/} share nice ideas with the
above literature, trying at the same time to avoid some typical
problems. For instance, both theories are explicitly covariant with
respect to changes of coordinates, even time--dependent ones. This
feature partially comes from the cosymplectic structure of the phase
space in the classical theory, which is also the classical background
for the quantum theory, and overcomes the problem of explicit time
independence of symplectic mechanics. On the other hand, the
cosymplectic structure of {\em CCG\/} is of a special, physically
reasonable, type. Therefore, it excludes those problems of the general
cosymplectic formalism, which have no physical meaning. Additionally,
both theories are covariant with respect to the choice of units of
measurement, because of the fact that the geometric framework
incorporates `unit spaces'. This kind of covariance requires different
geometric techniques compared to the standard literature. The theory
provides a model for classical and quantum holonomically constrained
systems of particles, supported by non trivial physically relevant
examples, such as the quantised rigid body \cite{MTV99}. 

We think that these aspects are promising for the theory to be a
mathematical framework for quantum mechanics.
  
\myskip

The goal of this paper is to analyse systematically the symmetries
of the structures involved in the covariant classical Galilei theory
and to introduce an associated momentum map. In a subsequent paper
we shall apply these results to the covariant quantum Galilei theory. 

\myskip

In the second section we summarise the basic aspects of the
covariant classical Galilei theory 
\cite{JadJanMod98,JadMod93,Mod99,MoVi95}
emphasising the natural bijections between the fundamental 
geometric objects of the theory.

In the model, classical spacetime is an $(n+1)$--dimensional 
manifold fibred over a 1-dimensional affine space. Spacetime is
supposed to be equipped with further geometric objects. Namely,
a scaled vertical metric of spacetime, called spacelike metric,
a linear connection of the tangent bundle of spacetime which 
preserves the time form, called spacetime connection
({\em gravitational field}), and a scaled 2--form of spacetime
({\em electromagnetic field}).   
The phase space is taken to be the first jet space with respect to 
the spacetime fibring.  

We shall see that there is a natural bijective correspondence between
a spacetime connection, a distinguished affine connection
of the phase space, called phase connection, and a distinguished 
(nonlinear) homogenous connection of the phase space, called
dynamical connection.
Moreover, there is a natural bijective correspondence between 
pairs of a phase connection and a spacelike metric and 
distinguished 2--forms of the phase space, called dynamical 
phase 2--forms. On the other hand, there is a natural bijective 
correspondence between pairs of a dynamical connection and a 
spacelike metric and distinguished 2--forms of the second jet space, 
called horizontal phase 2--forms. 

By means of the natural correspondences, we can regard the 
gravitational field as a dynamical phase 2--form of phase space if
a spacelike metric is chosen. 
Then, the electromagnetic field can be incorporated in the 
gravitational objects through a `minimal coupling'.
This yields a total dynamical phase 2--form. Assuming the closure 
of the gravitational and the electromagnetic 2--forms as the
dynamical equations for these fields, the postulated structure yields
naturally a cosymplectic structure on phase space.  
The dynamical connection that corresponds to the total dynamical
phase 2--form turns out to be the (scaled) Reeb vector field for 
this cosymplectic structure 
\cite{Albert89,CdLL92,dLS93,LiMa87,Mar83}.  
Consequently, it yields naturally the dynamics 
on spacetime, associated with the total dynamical
phase 2--form, as the geodesic flow of the corresponding dynamical 
connection. This leads directly to the notion of conserved quantities.

Equivalently, the dynamics can be described using the contact 
structure of the second jet space which yields naturally 
the (intrinsic) Euler--Lagrange morphism as the horizontal 
phase 2--form. 

On the other hand, any closed dynamical phase 2--form has local 
potentials. The spacetime-horizontal local potentials are said to be 
dynamical phase 1--forms and yield a time-horizontal part, 
called Lagrangian, according to the contact splitting induced by 
phase space. This turns out to be a (local) Lagrangian for the 
above (global and intrinsic) Euler--Lagrange morphism.
For any Lagrangian, its associated dynamical phase 1--form turns 
out to be the Poincar\'e--Cartan form for this Lagrangian.
Furthermore, given an observer, \ie any connection of spacetime,
a Poincar\'e--Cartan form splits into the observed Hamiltonian
and the observed momentum.

Finally, we recall the notion of $\tau$-Hamiltonian lift 
\cite{JadJanMod98,JadMod93,Mod99}, which is a covariant lift
of functions of the phase space to vector fields of phase space
motivated from the covariant quantum Galilei theory. 
We will see that this lift plays an important role in the 
definition of momentum map. 
  
\myskip

In the third section, we analyse systematically the infinitesimal
symmetries of the geometric objects that we have introduced so far. 
Hereby, we are mainly interested in holonomic symmetries.
Natural bijections are used to find the relations between the symmetries of the different objects. 
 
After recalling the basic properties about symmetries and 
infinitesimal symmetries we apply this notion to the
objects {\em CCG\/}.
However, some of the geometric objects that we need for our analysis
are not tensors. Therefore, we show first, 
how to get well defined expressions for the infinitesimal symmetries 
of these objects. 
We need these results to show in Theorem \ref{KGamgam} 
that any holonomic infinitesimal symmetry of a spacetime connection 
is a symmetry of the corresponding phase connection, and vice-versa.
Moreover, any holonomic infinitesimal symmetry of a spacetime 
connection turns out to be a symmetry of the corresponding dynamical 
connection, and vice--versa.
On the other hand, Theorem \ref{OmeGGam} shows that any holonomic 
infinitesimal symmetry of a dynamical phase 2--form is also a 
symmetry of the corresponding spacelike metric and the corresponding 
affine connection of the phase space, and vice--versa. 
Similarly, we prove in Theorem \ref{OmeE} that any holonomic 
infinitesimal symmetry of the Euler--Lagrange morphism is also a 
symmetry of the corresponding pair of a spacelike metric and a 
dynamical connection, and vice--versa. 
This yields directly the equivalence of (holonomic) symmetries
of the dynamical phase 2--form and the corresponding Euler-Lagrange 
morphism. 

In subsection \ref{symNoether}, we restrict our attention
to closed dynamical 2--forms.
We see, how conserved quantities are related to symmetries of such
forms. This yields directly the Theorem of Noether \ref{Thepotential} 
for symmetries of Poincar\'e--Cartan forms. 
Using Theorem \ref{symTheL} which is also suggested from a natural 
bijective correspondence and which shows that any holonomic 
infinitesimal symmetry of a Poincar\'e--Cartan form is a symmetry  
of the corresponding Lagrangian, and vice versa, we get another
equivalent version of the Noether theorem in Corollary 
\ref{Lpotential} which may be
more popular to the physical reader.  

In subsection \ref{momentummap}, we provide a definition of 
covariant momentum map for an action of a group of symmetries 
of the cosymplectic structure in our model.
This definition is similar to momentum map in presymplectic
and cosymplectic literature \cite{Albert89,CdLL92,dLS93,LiMa87,Mar83}.
However, the covariance of our theory requires the concept of
$\tau$-Hamiltonian lifts. This leads us directly to the quantisable 
functions and thus to the quantum theory. 

We see that we can associate to any infinitesimal generator, 
associated with the infinitesimal action, a pair, namely a conserved 
quantity and an element of the vector space of time units. 
Such a pair is determined up to an additive real constant
of the conserved quantity.  
We say a map that associates to each element of the Lie algebra 
for the group of cosymplectic symmetries such a pair to be a 
momentum map for the (infinitesimal) action of the group.  
We prove that the $\tau$-Hamiltonian lift of any component of 
a momentum map is equal to the infinitesimal generator associated 
with the group of symmetries for which this component was defined. 
Moreover, we introduce, a bracket for pairs such that the map
that associates to a pair its $\tau$-Hamiltonian turns out to
be a morphism of Lie algebras.
        
For the case of an action that, additionally, projects on an action
of the group on spacetime, we find that the components of 
a momentum map are of special type. Namely, they are 
`quantisable functions', \ie functions which are polynomials of
second degree in the velocities, and whose second derivative (with
respect to velocities) is proportional to the metric. Such functions
are of fundamental importance  in the quantum part of this model
\cite{JadJanMod98,JadMod93,Mod99}. 
This feature is new with respect to standard symplectic momentum map formalism, and is mainly due to covariance requirements. Its importance is fundamental in quantum mechanics.
In this case, a momentum map is determined by the first component
of the pairs. Then, we consider an action of a group of symmetries 
that, additionally, is a group of symmetries of a Poincar\'e--Cartan 
form. Here, the momentum map turns out to be 
unique and it is determined by the Poincar\'e--Cartan form.   

Finally, three very simple examples are provided in order to show the 
machinery at work. These examples show also, that in standard 
time--independent situations of particle mechanics we get the same
results as standard symplectic mechanics \cite{MaRa95}. However, 
we are able to treat time--dependent cases, as well. 
Therefore, we can relate time translations to the Hamiltonian in the 
standard (relativistic) way. For a less trivial  application of the 
momentum map in our theory, see \cite{MTV99}. 

\myskip

We find our results promising also in view of a next research about 
symmetries of quantised systems.

\myskip

{\bf Acknowledgements}.

This research has been supported by LGFG, DAAD, GNFM, MURST, 
and the Universities of Florence, Lecce and Mannheim.

We thank E. Binz, M. Modugno, C. Tejero Prieto for stimulating 
discussions.
\section{Covariant classical Galilei theory}
\label{Covariant classical Galilei theory}
\setcounter{assump}{3}

In order to be rather self contained, we start by recalling the main
points of the covariant classical Galilei theory. 
\cite{CJM95,JadMod93,JadJanMod98,Mod99}
\subsection{Unit spaces}

Now, we are going to assume the fundamental spaces of units of
measurement and coupling constants.

The theory of {\em unit spaces\/} has been developed in 
\cite{JadJanMod98,JadMod93} to make the independence of classical and 
quantum mechanics from scales explicit. Unit spaces are defined 
similarly to vector spaces, but using the abelian semigroup $\R^{+}$ 
instead of the field of real numbers $\R$. In particular, 
{\em positive\/} unit spaces are defined to be $1$--dimensional
(over $\R^{+}$) unit spaces. It is possible to define $n$--th 
tensorial powers and $n$--th roots of unit spaces. Moreover, if 
$\BP$ is a positive unit space and $p \in \BP$, then we denote by 
$1/p \in \BP^*$ the dual element. Hence, we can set 
$\BP^{-1} \byd \BP^{*}$. In this way, we can introduce rational 
powers of unit spaces.

We assume the following unit spaces.

--$\BT$ , the oriented one--dimensional semi-vector space of {\em time
intervals\/};

--$\BL$ , the positive unit space of {\em length units};

--$\BM$ , the positive unit space of {\em mass units};

We denote by $\olin{\BP} \byd \BP \otimes \BR$ the associated vector
space to the unit space $\BP$.  An element $u_{0}\in\BT$ (or
$u^{0}\in \BT^{-1}$) represents a 
{\em time unit of measurement\/}, a {\em  charge\/} is represented by
 an element 
$q \in \BQ \byd \BT^{-1} \otimes \BL^{3/2} \otimes \BM^{1/2}$, and
 a {\em particle\/} is represented by a pair $(m,q)$,  where $m$ is a
 mass and $q$ is a charge. A tensor field with values into mixed 
rational powers of $\BT$, $\BL$, $\BM$ is said to be {\em scaled\/}. 
We assume the {\em Planck's  constant\/} 
$\hbar\in\BT^{-1}\ten\BL^2\ten\BM$.

We will often be involved with the Lie derivative of scaled tensor 
fields. It can be shown \cite{JadJanMod98,JadMod93} that the
Lie derivative commutes with the scaling.
 
In the following, we assume all manifolds and maps to be $\cin$.

\subsection{Spacetime and phase space}

\bAs\label{spacetime}
We assume {\em spacetime\/} to be an $(n+1)$-dimensional oriented 
fibred manifold
\beq
t : \bE \to {\bT}
\eeq
over a 1-dimensional oriented affine space ${\bT}$ ({\em time\/})
associated with the vector space $\bar{\BT}$, where $n \in \BN$ with 
$n \ge 2$.$\END$
\eAs

\myskip

We shall refer to {\em spacetime charts\/} $(x^0, x^i)$, which are
adapted to the fibring, to a time unit of measurement $u_0 \in \BT$
and to the chosen orientation of $\bE$. The index $0$ will refer to 
the base space, Latin indices $i,j, \dots = 1,2,3$ will refer to the 
fibres, while Greek indices $\lam, \mu, \dots = 0, 1, 2, 3$ will 
refer both to the base space and the fibres.

A {\em motion\/} is defined to be a section $s:\bT \to \bE$. The
coordinate expression of a motion $s$ is of the type $s^i \byd x^i
\com s : \bT \to \R$.

We shall be involved with the {\em tangent bundle\/} 
$\tau_{\bE} : T\bE \to \bE$ and the 
{\em vertical tangent subspace\/} 
$V\bE \byd \ker Tt \sub T\bE$. 
We denote the charts induced on $T\bE$ by 
$(x^\lam, \dot{x}^\lam)$; moreover, we denote the induced local bases
of vector fields, of forms and of vertical forms of $\bE$,
respectively, by
$(\der_\lam)$, $(d^\lam)$ and $(\check{d}^i)$.
\myskip

The {\em phase space\/} is defined to be the first jet space of 
motions $t^1_0 : J_1\bE \to \bE$.
We denote the charts induced on $J_1\bE$ by $(x^0,x^i,x^i_0)$.
We will be involved with the second jet space of motions
$t^2_0 : J_2\bE \to \bE$.  
We denote the charts induced on $J_2\bE$ by 
$(x^0,x^i,x^i_0, x^i_{00})$.

The velocity of a motion $s$ is the section 
$j_1 s : \bT \to J_1 \bE$, with coordinate expression 
$x^i_0 \com j_1 s = \der_0 s^i$.

The phase space is equipped with the natural maps
$\D_1 : J_1\bE \to \BT^* \ten T\bE$
and
$\vartheta_1: J_1\bE \to T^*\BE \uten{\bE} V\bE$,
with coordinate expressions
$\D_1 = u^0 \otimes \D_{1,0} = u^0 \otimes 
(\der_0 + x^i_0 \, \der_i)$ and
$\vartheta_1= \vartheta_1^i \otimes \der_i = (d^i - x^i_0 \, d^0) 
\otimes \der_i$.
Analogously, the jet space $J_2\bE$ is equipped with the natural
maps
$\D_2 : J_2\bE \to \BT^* \otimes TJ_1\bE$
and
$\vartheta_2 : J_2\bE \to T^*\BE \uten{J_1\bE} VJ_1\bE$,
with coordinate expressions
$\D_2 = u^0 \otimes \D_{2,0} = u^0 \otimes (\der_0 + x^i_0 \, 
\der_i + x^i_{00} \, \der_i^0)$
and
$\vartheta_2 = \vartheta_2^i \otimes \der_i = 
(d^i_0 - x^i_{00} \, d^0)
\otimes \der^0_i + (d^i - x^i_0 \, d^0) \otimes \der_i$.

An {\em observer\/} is defined to be a section
$o : \bE \to J_1\bE$. Its coordinate expression is of the type
$o = u^0 \ten (\der_0 + o^i_0 \, \der_i)$, 
where
$o^i_0 : \bE \to \R$.

An observer $o$ can be regarded as a scaled vector field of $\bE$.
The {\em integral motions\/} of an observer $o$ are defined to be the
motions $s$ such that $j_1 s = o \com s$. An observer $o$ yields 
locally a fibred splitting $\bE \to \bT \car \bP$, where $\bP$ is 
the manifold of integral motions of $o$. An observer is said to be 
{\em complete\/} if it yields a global splitting of $\bE$. A spacetime
chart is said to be {\em adapted\/} to $o$ if it is adapted to the
local splitting of $\bE$ induced by $o$, \ie if $o^i_0 = 0$. 

An observer $o$ can be regarded as a connection of the fibred
manifold $\bE \to \bT$. Accordingly, it yields the translation fibred 
isomorphism
$\nab[o] : J_1\bE \to \BT^*\ten V\bE$, given by
$\nab [o](e_1) \byd e_1 - o(t^1_0(e_1))$.
We have the coordinate expression
$\nab[o] = (x^i_{0}-o^i_{0}) \, d^0\ten\der_{i}$.\\[2mm]

\subsection{Natural bijective correspondences}

In the following we define distinguished objects living on 
spacetime or phase space and we investigate their relations. 
In a concrete model of a classical system these objects will be 
either determined by further assumptions or as a consequence of the assumptions.
    
\bDf
A scaled vertical Riemannian metric
\bEq
g: \bE \to \BL^2 \ten (V^*\bE \uten{\bE} V^*\bE) \,
\eEq
is said to be a {\em spacelike metric}.\END
\eDf

The coordinate expression of a spacelike metric $g$ is
$g = g_{ij} \, \Check{d}^i \otimes \Check{d}^j$,
where $g_{ij} : \bE \to \bar{\BL}^2$.

Given a mass $m \in \BM$, it is convenient to introduce a
``normalised'' metric $G \equiv \frac{m}{\h} g$, with coordinate 
expression $G = G^0_{ij} \, u_0 \ten \Check{d}^i \ten \Check{d}^j$,
where $G^0_{ij} : \bE \to \R$.

A metric $G$ yields a family of Riemannian connections of the 
fibres of $\bE \to \bT$
\bEq\label{kap}
\varkap : V\bE \to V^*\bE \uten{V\bE} VV\bE
\eEq
with coordinate expression
$\varkap = d^k\ten(\der_k + \varkap_k{}^i{}_h \, \dot{x}^h \,
\dot\der_i)$, where $\varkap_k{}^i{}_h$ are the usual Christoffel 
symbols on the fibres of $\bE \to \bT$ related to $G$.

Next, we analyse distinguished connections that can be defined on
spacetime or phase space. We recall the fact, that for any fibred 
manifold $p: \bF \to \bB$, there are natural bijective 
correspondences 
\beq
c \leftrightarrow \nu  \leftrightarrow \nab
\eeq
between the {\em connections} 
$c: \bF \to T^*\bB \uten{\bF} T\bF$, 
the {\em vertical projectors} 
$\nu: \bF \to T^*\bF \uten{\bF} V\bF$ 
and the {\em covariant derivatives} 
$\nab: J_1\bF \to T^*\bB \uten{\bF} V\bF$ of $p$.
In order to give their coordinate expressions, we take, only in
this  case, greek indices for the basis coordinates and
latin indices for the fibres. Then, their expressions are
$c = d^\lam \ten (\der_\lam + c^i_\lam \der_i)$,
$\nu[c] = (d^i - c^i_{\lam} d^\lam) \ten \der_i$
and $\nab[c] = (x^i_\lam - c^i_\lam) d^\lam \ten \der_i$.  

\bDf
A {\em spacetime connection\/} is defined to be a $dt$--preserving
torsion free linear connection of the vector bundle $T\bE \to \bE$
\bEq
K : T\bE \to T^*\bE \uten{T\bE} TT\bE \,.\END
\eEq 
\eDf
The coordinate expression of any spacetime connection $K$ is of the 
type $K = d^\lam \ten  (\der_\lam + K_\lam{}^i{}_\nu \, 
\dot{x}^\nu \, \dot\der_i)$, 
where 
$K_\nu{}^i{}_\lam = K_\lam{}^i{}_\nu : \bE \to \R$.
The compatibility with $dt$, 
\ie the condition $\nab[K] dt = 0$, is expressed by 
$K_\mu{}^0{}_\nu = 0$.

The restriction of a spacetime connection $K$ to the vertical tangent
bundle is a linear connection
$K' : V\bE \to T^*\bE \uten{V\bE} TV\bE$. 

A spacetime connection $K$ is said to be {\em metric\/} if
$\nab[K'] \, G = 0$. In this case, $K$ is partially determined
by the metric according to the local formulas
$K_{ihj} = 
- \tfrac12 (\der_i G_{hj} + \der_j G_{hi} - \der_h G_{ij})$ and 
$K_{0ij} + K_{0ji} = - \der_0 G_{ij}$
where indices have been raised or lowered by the metric G.

\bDf
A {\em phase connection\/} is defined to be a torsion--free affine
connection of the affine bundle $J_1\bE \to \bE$
\bEq
\Gam : \bE \to T^*\bE \uten{J_1\bE} TJ_1\bE \,.
\eEq
Here the torsion is defined through the vertical valued form 
$\vartheta_1$ \cite{Mod91}.\END 
\eDf
The coordinate expression of $\Gam$ is of the type
$\Gam = d^\lam \ten 
\big(\der_\lam + \Gam_\lam{}^i_0 \, \der^0_i\big)$, 
where
$\Gam_\lam{}^i_0 \equiv \Gam_\lam{}^i_0{}^0_h \, x^h_0 +
\Gam_\lam{}^i_0{}^0_0$
and
$\Gam_\lam{}^i_0{}^0_j, \; \Gam_\lam{}^i_0{}^0_0 : \bE \to \R$.

It can be easily seen in coordinates, 
that there is a natural bijective correspondence 
\cite{JadMod93,Jan94,Mod99}
\beq
K \leftrightarrow \Gam[K]
\eeq 
between spacetime connections and phase connections
with coordinate expression 
$\Gam_\lam{}^i_0{}^0_j = K_\lam{}^i{}_j$ and
$\Gam_\lam{}^i_0{}^0_0 = K_\lam{}^i{}_0$.

A {\em second order connection\/} of spacetime is defined to be 
a (nonlinear) connection of the fibred manifold $J_1\bE \to \bT$
\bEq
\gam : J_1\bE \to \BT^* \ten TJ_1\bE \,,
\eEq
which is projectable on the contact map $\D_1$.
The coordinate expression of $\gam$ is of the type 
$\gam = u^0 \ten (\der_0 + x^i_0 \der_i + \gam_0{}^i_0)$,
where $\gam_0{}^i_0 : J_1\bE \to \R$. 
\bDf
A second order connection is said to be a 
{\em dynamical connection} if it is 
``homogeneous'' in the sense of \cite{KolMod90}, \ie if
its coordinate expression is of the type 
$\gam_0{}^i_0 \equiv \gam_0{}^i_0{}^0_h{}^0_k \, x^h_0 x^k_0 + 
2 \, \gam_0{}^i_0{}^0_h \, x^h_0 +
\gam_0{}^i_0{}^0_0$ and
$\gam_0{}^i_0{}^0_h{}^0_k, \; \gam_0{}^i_0{}^0_h, \;
\gam_0{}^i_0{}^0_0 : \bE \to \R$.\END
\eDf 

We will be involved with the covariant derivative of a dynamical 
connection, \ie the morphism
$\nab[\gam]: J_2\bE \to \BT^* \ten \BT^* \ten V\bE$ with 
coordinate expression 
$\nab[\gam] = 
u^0 \ten (x^i_{00} - \gam^i_{00}) d^0 \ten \der_i$. 

We can easily see in coordinates that the map 
$\Gam \mto \gam[\Gam] \byd \D_1 \con \Gam$
is a natural bijective correspondence \cite{MoVi95}
\beq
\Gam \leftrightarrow \gam[\Gam]
\eeq
between the phase connections 
and dynamical connections, namely such that 
$\gam_0{}^i_0{}^0_h{}^0_k = \Gam_h{}^i_0{}^0_k$,
$\gam_0{}^i_0{}^0_h  = \Gam_h{}^i_0{}^0_0$ and
$\gam_0{}^i_0{}^0_0  = \Gam_0{}^i_0{}^0_0$.

Next, we define distinguished forms of $J_1\bE$ and $J_2\bE$
that can be defined through the above objects.
\bDf
A 2--form $\Ome$ of $J_1\bE$ of the type
\bEq \label{Omega}
\Ome[G, \Gam] =
\nu[\Gam] \, \bar\wed \, \vartheta_1: J_1\bE \to \Lam^2T^*J_1\bE \,,
\eEq
where $\nu[\Gam]$ is the vertical projection associated with a 
phase connection $\Gam$ and where the contracted wedge product 
is taken with respect to a spacelike metric $G$
a {\em dynamical phase 2--form}.\END 
\eDf
We have the coordinate expression
$\Ome[G, \Gam] = G^0_{ij} \, (d^i_0 - \Gam_\lam{}^i_0{} \, d^\lam) 
\wed \vartheta_1^j 
= G^0_{ij} \, (d^i_0 - \gam_0{}^i_0 d^0 - \Gam_h{}^i_0 \, 
\vartheta_1^h) \wed \vartheta_1^j$.

We observe that the above form is the only natural $2$--form which can be obtained from $\Gam$ and $G$ \cite{Jan94,JadMod93}. 
Moreover, the form $\Ome[G, \Gam]$ is non degenerate in the sense that $dt \wed \Ome[G, \Gam]^n$ is a volume form of $J_1\bE$.
We can easily prove that there is a unique scaled vector field
$X: J_1\bE \to TJ_1\bE$ such that $i_X\, dt = 1$ and 
$i_X\, \Ome[G, \Gam] = 0$, namely, 
$X = \gam[\Gam]$, the dynamical connection. 
This yields a natural bijective correspondence 
\beq
(G, \Gam) \leftrightarrow \Ome[G, \Gam]
\eeq
between the pairs of a spacelike metric and a phase connection 
and the dynamical phase 2--forms.

\bDf\label{Euler}
A 2--form $\cE$ of $J_2\bE$ of the type
\bEq \label{EulerLagrange}
\cE[G, \gam] =
\nab[\gam] \, \bar\wed \, \vartheta_1: J_2\bE \to \Lam^2T^*\bE \,,
\eEq
where $\nab[\gam]$ is the covariant differential associated with a
dynamical connection $\gam$ and where the contracted wedge product 
is taken with respect to a spacelike metric $G$ is called 
a {\em horizontal phase 2--form}.\END
\eDf
We have the coordinate expression
$\cE[G,\gam] = G_{ij}^0 (x^i_{00} - \gam^i_{00}) d^0 
\wed (d^j - x^j_0 d^0)$.

Analogously to the case of a dynamical phase 2--form, it turns out
that there is a natural bijective correspondence  
\beq
(G, \gam) \leftrightarrow \cE[G, \gam]
\eeq
between the pairs of a spacelike metric and a dynamical connection 
and the horizontal phase 2--forms.

Consequently, we obtain a natural bijective correspondence 
\beq
\Ome[\Gam, G] \leftrightarrow \cE[\nab\gam[\Gam], G]
\eeq 
between the dynamical phase 2--forms and the horizontal phase 
2--forms.

It turns out that the horizontal phase 2--form $\cE$ corresponding to 
a dynamical phase 2--form $\Ome$ coincides with the horizontal part 
of $\Ome$ according to the contact splitting of forms of $J_1\bE$
induced by $J_2\bE$.

Let us consider the case of a closed dynamical phase 2--form $\Ome$.
We can prove \cite{JadJanMod98,JadMod93,Mod91,Mod99} that 
$d\Ome = 0$ is equivalent to the conditions that
$\nab[K']G = 0$ and that, in coordinates, the curvature
$R[K]: \bE \to \Lam^2 T^*\bE \uten{\bE} V\bE \uten{\bE} T^*\bE$
fulfills $R_\lam{}^i{}_\mu{}^j = R_\mu{}^j{}_\lam{}^i$.
Given an observer $o$ we can define the 2--form of $\bE\,$ 
$\Phi[o] \byd 2o^*\Ome: \bE \to \owed{2} T^*\bE$.
It turns out that $d\Ome = 0$ is also equivalent to the conditions
$\nab[K']G = 0$ and $d\Phi[o] = 0$. 

\bRm
If the phase space $J_1\bE$ is equipped with a closed dynamical 
phase 2--form $\Ome$, the triple 
$(J_1\bE, \Ome, dt)$ turns out to be a cosymplectic manifold.
The corresponding dynamical connection turns out to be the (scaled) 
Reeb vector field for this cosymplectic structure.
\eRm

Moreover, a closed dynamical phase 2--form $\Ome$ admits potential 
1--forms of $J_1\bE$. In the following we introduce a special kind 
of such potential forms. 
\bDf
A horizontal 1--form $\The: J_1\bE \to T^*\bE$ such that 
$ d \The = \Ome$ where $\Ome$ is a closed dynamical phase 2--form
is said to be a {\em dynamical phase 1--form} associated with $\Ome$.
\END
\eDf
For any observer $o$, the expression of a dynamical phase 1--form 
associated with $\Ome$ in adapted coordinates is given by
$\The = - (\frac12 G^0_{ij} x^i_0 x^j_0 - A_0) d^0
        + (G^0_{ij} + A_i) d^i$ 
where $A_\lam d^\lam$ is a potential of the closed 2--form 
$\Phi[o] = 2 o^* \Ome$.
Clearly, a dynamical phase 1--form associated with $\Ome$ is 
determined up to a closed 1--form of $\bE$. 

According to the contact splitting of $\The$ induced by $J_1\bE$
we define the following
\bDf
The {\em Lagrangian\/} $\cL[\The]$ associated with a dynamical phase 
1--form $\The$ is defined to be the time-horizontal 1--form
\bEq
\cL \byd \D_1 \con \The : J_1\bE \to T^*\bT\,.
\eEq
The {\em momentum\/} $\cP[\cL]$ of the Lagrangian $\cL$ is defined 
to be the vertical derivative of $\cL$ with respect to the fibring  
$t^1_0: J_1\bE \to \bE$, \ie $\cP \byd V_{\bE} \cL$.\END 
\eDf
We have the coordinate expressions
$\;\cL[\The] = (\frac12 G^0_{ij} x^i_0 x^j_0 + A_i x^i_0 + A_0) d^0\,$
and
$\;\cP[\cL] = (G^0_{ij} x^j_0 + A_i) (d^i - x^i_0 d^0)\,$.  
  
It turns out that $\The$ splits into 
$\The = \cL[\The] + \cP[\cL[\The]]$. This splitting coincides with the
contact splitting of $\The$ induced by $J_1\bE$.
 
This yields directly the natural bijective correspondence
\beq
\The \leftrightarrow \cL[\The]
\eeq 
between dynamical phase 1--forms and Lagrangians.

Moreover, $\The[\cL]$ turns out to be the Poincar\'e--Cartan form
associated with the Lagrangian $\cL$. Hence, in the following, we 
say any dynamical phase 1--form to be a Poincar\'e--Cartan form.
We stress the fact that the objects $\The$, $\cL$ and $\cP$ do
not depend on an observer but on a chosen local gauge. It is for 
practical convenience that we have given the coordinate expressions 
with respect to an observer.

Given a closed dynamical phase 2--form $\Ome$ it can be proved that 
the horizontal phase 2--form $\cE[\Ome]$ coincides with the 
Euler-Lagrange morphism associated with any Lagrangian $\cL[\The]$
where $\The$ is a dynamical phase 1--form associated with $\Ome$.
Hence, in the following, we say a horizontal phase 2--form to be an
Euler-Lagrange morphism, when we are dealing with closed dynamical 
phase 2--forms. 

The above results concerning $\The$, $\Ome$, $\cL$ and $\cE$
are described by the following commutative diagram
$$
\begin{CD}
\The @>d>> \Ome
\\
@V{i_{\D_1}}VV @VV{i_{\vartheta_2}i_{\D_2}}V
\\
\cL @>\eps>> \cE   
\end{CD}
$$
This diagram can be regarded as a piece of a more comprehensive 
natural bicomplex, which accounts for the Lagrangian formalism via
a cohomological scheme \cite{Kru90,MoVi95,Vit98}. 

We summarise the above results in the following proposition
\bPr
The following natural bijective correspondences hold
\bEq
\label{correspondenceKGamgam}
&K \leftrightarrow \Gam \leftrightarrow \gam
\\
\label{correspondenceOmeE}
&\Ome \leftrightarrow (\Gam, G) \leftrightarrow (\gam, G) 
\leftrightarrow \cE
\\
\label{correspondenceTheL}
&\The \leftrightarrow \cL
\eEq
\ePr

It can be seen that there is another splitting of a 
Poincar\'e--Cartan form $\The$ which is observer dependent. 
Namely, given an observer $o$, each Poincar\'e--Cartan 
form splits, according to the splitting of $T^*\bE$ induced by
$o$ into the morphism 
$- \cH[o,\The] \byd - o \con \The: J_1\bE \to T^*\bT\,$,
called observed {\em Hamiltonian}, and the morphism
$\cP[o,\The] \byd \vartheta_1\con (\nu[o] \con \The): 
J_1\bE \to T^*\bE\,$,
called observed {\em momentum}.

\subsection{Classical dynamics}
\label{dynamics}

We are involved with two different approaches to the classical 
dynamics in our context. 
The first seems to be more direct and consists 
in defining directly the law of particle motion, namely,
the (generalised) Newton law.
\bDf\label{particle motion}
Let $\gam$ be a dynamical connection and $s$ a motion.
Then, the condition on $s$,
$$
\nab[\gam] j_{1}s \byd j_2s - \gam \com j_1s = 0\,. \eqno\END
$$ is said to be the 
{\em law of motion} for the dynamical connection $\gam$
\cite{CJM95,JadMod93,JadJanMod98,Mod99}. 
\eDf

The law of motion has the coordinate expression
$\der_0\der_0 s^i - \gam^i_{00} \com s = 0$.

A dynamical connection $\gam$ can be regarded (up to a time scale) 
as a vector field of $J_1\bE$, hence, a motion fulfilling the above 
equation is just an integral curve of $\gam$. 

It is easy to see that a motion $s$ fulfills the above law if and 
only if, for any $f : J_1\bE \to \R$, we have 
\bEq\label{Poisson}
d(f \com j_1s) = (\gam . f)\com j_1s\,.
\eEq
where $\gam . f \byd df(\gam)$.
In the particular case when $\gam.f = 0$ we call $f$ a 
{\em conserved quantity\/}.

\myskip

On the other hand, if the metric $G$ is given by a concrete model,
the natural bijective correspondence
$\cE \leftrightarrow (G, \gam)$ leads to an equivalent approach
to the equations of motion, namely, the equation
$\cE(\cL) j_2s = 0$ for any Lagrangian $\cL$. 
\subsection{Hamiltonian lift and quantisable functions}

Let us consider a dynamical phase 2--form $\Ome$. Then,
$\Ome$ yields in a natural way the Hamiltonian lift of functions 
$f: J_1\bE \to \R$ to vertical vector fields 
$H[f]: J_1\bE \to VJ_1\bE$. 
The musical morphism $\Ome\Fla: VJ_1\bE \to T_\gam^*J_1\bE$ 
turns out to be an isomorphism of vector spaces between vertical 
vector fields of $J_1\bE$ and forms of $J_1\bE$ that annihilate 
the corresponding $\gam$. 
Clearly, in the case when $\Ome = \Ome[G, \Gam]$, 
then $\gam = \gam[\Gam]$.  

More generally, taking into account the independence of units,
the choice of a time scale $\tau: J_1\bE \to T\bT$ yields,
in a natural (covariant) way, the $\tau$-Hamiltonian lift of functions
$f: J_1\bE \to \R$ to vector fields 
\bEq
H_\tau[f]:=  <\tau,\gam> +\, (\Ome\Fla)^{-1} (df - <\gam . f, dt>):
J_1\bE \to TJ_1\bE\,
\eEq
whose time component is $\tau$. 
We observe, that it is the scaling of the time form
which requires such a (covariant) lift, rather than the standard lift
in cosymplectic mechanics.
Its coordinate expression is
\beq
H_\tau[f] = \tau^0 
     (\der_0 + x^h_0 \der_h + \gam_0{}^i_0 \der^0_h) + G^0_{hk} 
     (-\der^0_kf \der_h + 
           (\der_kf + 
              (\Gam_k{}^l_0 - G^0_{kr} G^{ls}_0 \Gam_s{}^r_0)\,
            \der^0_lf)\,
         \der^0_h)
\eeq
where $\tau^0 \byd <\tau,u^0>$. 

In view of later developments in the quantum theory, it can be proved
\cite{JadMod93,JadJanMod98} that $H_\tau[f]$ is projectable on a 
vector field $X[f]: \bE \to T\bE$ if and only if the following 
conditions hold:\,
i)\, the function $f$ is quadratic with respect to the affine fibres 
of $J_1\bE \to \bE$ with second fibre derivative $f'' \ten G$, where 
$f'':\bE \to T\bT$ and\,ii) $\tau = f''$. A function of this kind is 
called {\em special quadratic} and is of the type
\beq
f = \frac12 f^0 G^0_{ij} x^i_0 x^j_0 + f^0_i x^i_0 + \of,,
\eeq
with $f^0, f^0_i, \of: \bE \to \R$.

Next, assume that $\Ome$ is closed. Then,
the $\tau$-Hamiltonian lift yields a Poisson bracket for functions of
$J_1\bE$, namely, the bracket 
$\{f,\,g\} \byd 
= i_{H_0[f]}\,i_{H_0[g]}\,\Ome$.
We observe that if $\tau, \sig$ are time scales, then
$\{f,\,g\} = i_{H_\tau[f]}\,i_{H_\sig[g]}\,\Ome$.
The vector space of special functions is not closed under the
Poisson bracket, but it turns out to be an $\R$-Lie algebra through
the natural {\em special bracket} 
$[f,\,g] = \{f,\,g\} + \gam(f'') . g - \gam(g'') . f$,
where $\{f,\,g\}$ is the Poisson bracket of the functions
$f$ and $g$. Of particular interest are such special functions whose
time component is a constant. They are  called 
{\em quantisable functions with constant time component}. 
 
It is easy to see that we can apply this bracket also onto
scaled functions. In particular, the Hamiltonian is a (scaled)
quantisable function with constant time component.
                
\section{Symmetries in covariant classical mechanics}

In the this section we want to introduce the notion of symmetry
to the objects which we have defined on spacetime and phase space.
We give theorems about the relation between the symmetries of these
objects. The natural correspondences turn out to be an indicator  
for symmetry relations. We see that these results can be directly
compared to standard results of Hamiltonian or Lagrangian mechanics. 
On the other hand, we think that our (non standard) approach to
these results within the covariant framework is promising, especially 
in view of applications to quantum theory ({\em CQG\/}).
  
\subsection{Symmetries and infinitesimal symmetries}

First, we want to recall the basic facts about symmetries,
groups of symmetries and infinitesimal symmetries of 
manifolds, fibred manifolds and tensors.
  
Let $\bM$ be a manifold. Then, we define a 
{\em symmetry of the manifold\/}  
$\bM$ to be a diffeomorphism $f: \bM \to \bM$.

The diffeomorphisms $f: \bM \to \bM$ constitute a group, 
called the {\em diffeomorphism group} ${\it Diff}(\bM)$, 
which operates on the manifold $\bM$ via the natural left action
$\Phi: {\it Diff}(\bM) \car \bM \to \bM: (f,m) \mto f(m)$. 
The group ${\it Diff}(\bM)$ is infinite dimensional, hence, it is
difficult to deal with the full group. But, in practice,
we are interested in finite dimensional subgroups with the 
structure of a Lie group.

On the other hand, we are often interested in an ``abstract''
Lie group $\bG$ which acts on the manifold $\bM$ through
a left action $\Phi: \bG \car \bM \to \bM$. The map $\Phi$
yields a group morphism (which may still be denoted by $\Phi$)
$\Phi: \bG \to {\it Diff}(\bM)$. This map needs not to be injective.
In the particular case when this map is injective we can 
identify the ``abstract'' group $\bG$ with the corresponding
subgroup of ${\it Diff}(\bM)$.

By taking the tangent prolongation of the action $\Phi$ with respect
to $\bG$, at the unit element $e \in \bG$, we obtain the linear 
fibred morphism over $\bM$, $\der\Phi: T_e\bG \car \bM \to T\bM$.  

By considering the natural identification of $T_e\bG$ with the
Lie algebra $\Fg$ of left invariant vector fields of
$\bG$ we can write the above morphism as a map
\beq
\der\Phi: \Fg \to Sec(T\bM): \xi \mto X[\xi] 
                                    \byd \der\Phi(\xi)\,,
\eeq
which turns out to be an antihomomorphism of Lie algebras.
We call $\der\Phi: \Fg \car \bM \to T\bM$ an
{\em infinitesimal left action} of the Lie algebra $\Fg$
on $\bM$ and the vector field $X[\xi]$ of $\bM$  
{\em  infinitesimal generator\/} of the infinitesimal action 
corresponding to $\xi$. 

Clearly, if $\Phi: \bG \to {\it Diff}(\bM)$ is injective, then
also $\der\Phi: \Fg \to Sec(T\bM)$ is injective.
                                                         
The above discussion suggests the following definition.
We call a vector field $X:\bM \to T\bM$ an 
{\em infinitesimal symmetry of the manifold\/} $\bM$ 
(since its local flow is a local group of diffeomorphisms).
Now, by considering a left action $\Phi$ of a Lie group $\bG$ on
$\bM$, the set of infinitesimal generators  
$\{X[\xi]: \bM \to T\bM,\, \A\, \xi \in \Fg\}$ 
turns out to be a subalgebra of the Lie algebra of 
infinitesimal symmetries of $\bM$.

Now we extend the definition of symmetries to manifolds that
are equipped with further structure.

Let $p:\bE \to \bB$ be a fibred manifold. Then, we define a 
{\em symmetry of the fibred manifold\/} 
$p$ to be a fibred diffeomorphism $f$ of $\bE$ over $\bB$, \ie a 
symmetry $f:\bE \to \bE$ of the manifold $\bE$ which projects on a 
symmetry $\uf: \bB \to \bB$ of the base space
$\bB$.
Given a (Lie) group of symmetries of $\bM$ any  
infinitesimal generator $X[\xi]$ turns out to be a vector 
field $X: \bE \to T\bE$ which projects on a vector field 
$\uX:\bB \to T\bB$. This suggests to call any vector field
$X$ of $\bE$ that projects on a vector field $\uX$ of $\bB$ an
{\em infinitesimal symmetry of the fibred manifold\/} $p$
(since its local flow is a local group of fibred diffeomorphisms).

Let $f: \bM \to \bM$ be a symmetry of the manifold $\bM$. Then, the 
tangent prolongation $Tf: T\bM \to T\bM$ turns out to be a symmetry 
of the fibred manifold $\tau_{\bM}: T\bM \to \bM$.
Moreover, for each left action $\Phi: \bG \car \bM \to \bM$, 
the tangent prolongation 
$\olin{T}\Phi: \bG \car T\bM \to T\bM: (g, y) \mto T(\Phi_g)(y)$
turns out to be a left action, called the 
{\em tangent prolongation\/} of $\Phi$. 

Let $\nu$ be a tensor field of $\bM$, which is contravariant of order 
$s$ and covariant of order $r$.
Then, we define a {\em symmetry of the tensor field\/ } 
$\nu$ to be a diffeomorphism $f: \bM \to \bM$ such that 
$\nu \com \oten{s} Tf = \oten{r} Tf \com \nu$.   

As before, we can define groups of symmetries of $\nu$.
It turns out that each infinitesimal generator $X[\xi]$ associated 
with a (Lie) group of symmetries of the tensor field $\nu$ fulfills 
the equation $L_{X[\xi]}\, \nu = 0$. This suggests to call any
vector field $X$ of $\bM$ such that $L_X\,\nu = 0$ an   
{\em infinitesimal symmetry of the tensor field\/} $\nu$
(since its local flow is a local group of symmetries of the
tensor field).
 
Now, let $f: \bE \to \bE$ be a symmetry of the fibred manifold
$p: \bE \to \bB$. Then, for each $1 \leq k$, the $k$-jet
prolongation $J_k f: J_k \bE \to J_k \bE$ turns out to be a
symmetry of the fibred manifolds $p^k_h: J_k \bE \to J_h \bE$
and $p^k: J_k \bE \to \bB$, for each $1 \leq h < k$.
Moreover, for each left action of symmetries
$\Phi: \bG \car \bE \to \bE$, the $k$-jet prolongation
$\olin{J}_k \Phi: \bG \car J_k\bE \to J_k\bE:
                 (g, e_k) \mto J_k(\Phi_g)(e_k)$
turns out to be a left action, called the 
{\em $k$-jet prolongation\/} of $\Phi$.

Let us recall the natural involution $s: TT\bM \to TT\bM$
\cite{Godbi}. This map yields the natural prolongation
of each vector field $X: \bM \to T\bM$ to the vector field
$X_{(T)} \byd s \com TX: T\bM \to TT\bM$.
If $X = X^\lam \der_\lam$, then 
$X_{(T)} = X^\lam \der_\lam + \der_\mu X^\lam \dot{x}^\mu 
\dot{\der}_\lam$. The map $X \mto X_{(T)}$ turns out to be 
a morphism of Lie algebras.
Let $\Phi: \bG \car \bM \to \bM$ be a left
action of $\bG$ on $\bM$. Then, we obtain
$s \com \olin{T}\der\Phi = \der\olin{T}\Phi:
     \Fg \car T\bM \to TT\bM$,
hence, $s \com \olin{T}\der\Phi$ turns out to coincide with the
infinitesimal left action of the Lie algebra $\Fg$
on the manifold $T\bM$. 

We recall the natural map $r^k: J_kT\bE \to TJ_k\bE$ \cite{MaMo83}.
This map yields the natural prolongation of each vector field
$X: \bE \to T\bE$ to the vector field 
$X_{(k)} \byd r^k \com J_kX: J_k\bE \to TJ_k\bE$.
In particular, if $X = X^\lam \der_\lam + X^i \der_i$, then 
$X_{(1)} = X^\lam \der_\lam + X^i \der_i + 
 (\der_\mu X^i + \der_k X^i x^k_\mu - \der_\mu X^\lam x^i_\lam) 
 \der^\mu _i$.
The map $X \mto X_{(k)}$ turns out to be a morphism of Lie algebras.
Let $\Phi: \bG \car \bE \to \bE$ be a left
action of a group $\bG$ such that $\{\Phi_g , g \in \bG\}$
is a group of symmetries of the fibred manifold 
$p: \bE \to \bB$. Then, we obtain
$r^k \com \olin{J}_k\der\Phi = \der\olin{J}_k\Phi:
    \Fg \car J_k\bE \to TJ_k\bE$,
hence, $r^k \com \olin{J}_k\der\Phi$ turns out to coincide with the
infinitesimal left action of the Lie algebra $\Fg$
on the manifold $J_k\bE$.

In general, we say all above natural prolongation of symmetries, 
infinitesimal symmetries and actions to be {\em holonomic}.   
By abuse of language we often call the group $\bG$ a group of
symmetries if the left action $\Phi$ of $\bG$ is given such that
$\Phi_g$ is a symmetry for all $g \in \bG$.
 
\subsection{Infinitesimal symmetries of spacetime structures}
\label{contact}

Now, we apply the above general definitions of infinitesimal
symmetries to covariant classical mechanics.
However, some of the morphisms (contact maps, spacelike metric)
cannot be regarded as tensors, naturally. 
Consequently, a (direct) definition of their
symmetries would require naturality techniques \cite{KMS93}.
Instead, we show that it is possible in our case to give a meaning
to their infinitesimal symmetries keeping the standard Lie 
derivative.     

\subsubsection*{Symmetries of spacetime}

An infinitesimal symmetry of spacetime is an infinitesimal symmetry 
of the fibred manifold $t: \bE \to \bT$ which, additionally, 
preserves the affine structure of $\bT$. More precisely, we define an  
{\em infinitesimal symmetry of spacetime\/} as a vector field 
$X: \bE \to T\bE$ which is projectable on a vector field 
$\uX : \bT \to T\bT$ and such that $\uX$ is constant. 

An easy calculation shows that
\bPr\label{symspacetime}
A vector field $X: \bE \to T\bE$ is an infinitesimal symmetry 
of spacetime if and only if  $L_X\, dt = 0$.

If $X = X^0 \der_0 + X^i \der_i$ is the coordinate expression, 
then the conditions are equivalent to $\der_\mu X^0 = 0$.
\ePr
\bPf
In coordinates, any vector field $X$ of $\bE$ is of the type
$X = X^0 \der_0 + X^i \der_i$, where $X^0, X^i: \bE \to \R$
are functions. The time form writes as $dt = u_0 d^0$. 
Hence, $L_X dt = u_0 \ten (\der_0 X^0 d^0 + \der_i X^0 d^i)$.
Therefore, $L_X dt = 0$ if and only if $\der_i X^0 = 0$
and $\der_0 X^0 = 0$. But the first condition on $X^0$
means that the vector field $X$ is projectable and,
together with the second condition, that $X^0$ is even constant.

\vspace{-5mm}\rightline\QED
\ePf

\subsubsection*{Symmetries of the contact maps}

On any fibred manifold $p: \bE \to \bB$, the contact maps 
$\vartheta_1$ and $\D$ are not tensors. Hence, their symmetries
require naturality techniques. 
However, it is possible to use a standard Lie derivative 
for both objects by using the 1--dimensional affine structure 
of $\bT$ in the following way. 

The affine structure of $p^1_0: J_1\bE \to \bE$ yields the 
following Lemma
\bLm
The map $\vartheta_1$ can be naturally regarded as a (scaled) tensor
\beq
\vartheta_1: J_1\bE \to T^*\bE \uten{\bE} V\bE
    \hto \BT \ten (T^*J_1\bE \uten{J_1\bE} TJ_1\bE)\,.\END
\eeq
\eLm

An easy calculation gives the following lemma which relates 
infinitesimal spacetime symmetries and holonomic infinitesimal 
symmetries of $\vartheta_1$.
 
\bPr\label{vartheta}
Let $X$ be an infinitesimal spacetime symmetry.
Then, $X_{(1)}$ is a holonomic infinitesimal symmetry of $\vartheta_1$, 
\ie $L_{X_{(1)}}\,\vartheta_1= 0$.
\ePr
\bPf
If $X = X^0 \der_0 + X^i \der_i$ where 
$X^0 \in \R, X^i: \bE \to \R$, then,
$L_{X_{(1)}}\,\vartheta_1= u_0 \ten (\der_0 X^0 d^k \ten \der^0_k
  - \der_l X^0 x^k_0 d^l \ten \der^0_k 
  - \der_0 X^0 x^k_0 d^0 \ten \der^0_k) = 0$.

\vspace{-5mm}\rightline\QED 
\ePf
 
\bLm\label{symD}
Let $p: \bF \to \bB$ a fibred manifold. Let $X: \bF \to T\bF$
be a vector field which projects on a vector field 
$\uX: \bB \to T\bB$, and let $Y: \bF \to T\bB$ be a fibred morphism
(over the identity on $\bB$). 
Let $\tilde{Y}: \bF \to T\bF$ be any extension of $Y$, \ie
a vector field projectable on $Y$.
Then, $L_X\,Y \byd Tp \com (L_X\,\tilde{Y})$ is well defined,
\ie it does not depend on the extension $\tilde{Y}$ of  $Y$.

We obtain the following coordinate expression
\beq
L_X\,Y = (X^\mu \der_\mu Y^\lam - Y^\mu \der_\mu X^\lam 
           + X^i \der_i Y^\lam) \der_\lam\,.
\eeq
\eLm 
\bPf
In coordinates, we have
\beq
L_X\,\tilde{Y} &=& (X^\mu \der_\mu Y^\lam - Y^\mu \der_\mu X^\lam
            + X^i \der_i Y^\lam - \tilde{Y}^i \der_i X^\lam) \der_\lam
\\
       &&+ (X^\mu \der_\mu \tilde{Y}^j - Y^\mu \der_\mu X^j
            + X^i \der_i \tilde{Y}^j - \tilde{Y}^i \der_i X^j) 
            \der_j\,.
\eeq
Since $X$ is projectable, this yields
$Tp \com L_X\,\tilde{Y} = (X^\mu \der_\mu Y^\lam - 
 Y^\mu \der_\mu X^\lam + X^i \der_i Y^\lam)\der_\lam$.

This does not depend on the coefficients $Y_i$.
The result turns out to be independent of the chart.

\vspace{-5mm}\rightline\QED
\ePf
In the following we use Lemma \ref{symD} to define symmetries of $\D$.

\bPr\label{holsymD}
Let $X: \bE \to T\bE$ an infinitesimal spacetime symmetry.
Then, $X_{(1)}$ is a holonomic infinitesimal symmetry of $\D$, 
\ie $L_{X_{(1)}}\,\D_1= 0$.
\ePr
\bPf
$\D$ is a (scaled) fibred morphism over $\bE$. 
Thus, Lemma \ref{symD} yields the equality 
\beq
L_{X_{(1)}}\,\D_1= u^0 \ten (X^\lam \der_\lam \D^\mu_{1,0}
                    + X^i_0 \der^0_i \D^\mu_{1,0}
                    - \D^\lam_{1,0} \der_\lam X^\mu) \der_\mu
                = 0
\eeq
where the components of $\D_1$ were given by 
$\D^\mu_{1,0} = (1, x^i_0)$.  

\vspace{-5mm}\rightline\QED
\ePf

\subsubsection*{Symmetries of spacelike metrics}

Let $G$ be a spacelike metric. Clearly, $G$ is not a tensor 
of $\bE$. Hence, in order to define its infinitesimal symmetries, 
we use the following lemma
\cite{Mod99}.

\bLm\label{Lie derivative of a vertical covariant tensors}
Let us consider a fibred manifold $p: \bF \to \bB$, a vector field 
$X$ of $ \bF$ which is projectable on a vector field $\uX$ of 
$\bB$ and a vertical covariant tensor  
$\alp : \bF \to \oten{r} V^*\bF$. 
Then, the vertical restriction
$(L[X]\tilde{\alp})\,\Check{} : \bF \to \oten{r} V^*\bF$
of the Lie derivative $L[X]\tilde{\alp}$, where
$\tilde{\alp} : \bF \to T^*\bF$ is an extension of $\alp$, does
not depend on the choice of the extension $\tilde{\alp}$.

Hence, the Lie derivative
$L[X]\alp \byd (L[X]\tilde{\alp})\,\Check{} : \bF \to \oten{r} 
V^*\bF$
is well defined.

Its coordinate expression is
\beq
L_X\,\alp = (  X^\lam \der_\lam \alp_{j_1 ... j_r}
             + X^i \der_i \alp_{j_1 ... j_r}
            + \alp_{i j_2 ... j_r} \der_{j_1} X^i
            + \,... \,+ 
            \alp_{j_1 ... j_{r-1} i} \der_{j_r} X^i)
            d^{j_1} \ten ... \ten d^{j_r} 
\eeq
where $(j_1, \,...\,j_r)$ is any permutation of the fibre indices,
and where we have used greek indices for the coordinates of the
base space and latin indices for the fibres.
\eLm
\bPf
The coordinate expression shows that because of the projectabilty
of $X$ the vertical restriction does not contain coefficients of the 
type $\alp_{..\mu..}$. The result turns out to be independent of the 
chart.

\vspace{-5mm}\rightline\QED
\ePf

Thus, for each spacelike metric $G$, we call a projectable vector 
field $X$ of $\bE$ such that $L_X\,G = 0$ an 
{\em infinitesimal symmetry of the spacelike metric} $G$.
An easy calculation shows
\bPr\label{symexpressionG}
Let $X$ be an infinitesimal spacetime symmetry and $G$ a spacelike
metric. Then, we have the coordinate expression
\beq
L_X\,G = 
\{X^\lam (\der_\lam G^0_{ij})  + G^0_{kj} (\der_i X^k)
  + G^0_{ih} (\der_j X^k)\} d^i \ten d^j\,.\END
\eeq
\ePr

\subsubsection*{Symmetries of connections}

Let $K$ be a spacetime connection, $\Gam$ a phase connection
and $\gam$ a dynamical connection.
An easy calculation yields the following coordinate 
expressions.
\bPr\label{symexpressions}
Let $X$ be an infinitesimal spacetime symmetry. Then,
\beq
L_{X_{(1)}}\,\Gam &=&  
                \{ \der_\mu X^i_0 
               - (\Gam_{\lam}{}^i_{00} + 
                  \Gam_{\lam}{}^i_{0k} x^k_0)(\der_\mu X^\lam)
               - \der_\lam (\Gam_{\mu}{}^i_{00} +
                             \Gam_{\mu}{}^i_{0k} x^k_0) X^\lam
\\
             && - \Gam_{\mu}{}^i_{0k} X^k_0 
               + (\Gam_{\mu}{}^j_{00}
                  + \Gam_{\mu}{}^j_{0k} x^k_0)(\der_j X^i)\}          \, d^\mu \ten \der^0_i\,,
\\&&\\
L_{X_{(1)}}\,\gam &=& 
            u^0\{ X^0 (\der_0 \gam^i_{00}) 
                    + X^i (\der_j \gam^j_{00}) 
                    - (\der_0 X^i_0) 
                    - (\der_j X^i_0) x^j_0
\\                &&
                    + X^j_0 (\der^0_j \gam^i_{00}) 
                    - \gam^j_{00} (\der^0_j X^i_0) \}
\,\der^0_i\,, 
\\&&\\
L_{X_{(T)}}\,K &=& 
          \{(\der_\lam X^i) \Gam_{i}{}^k_{0\mu} \dot{x}^\nu
                          - \der_\lam \der_\nu X^k \dot{x}^\nu
        + X^\alp \der_\alp \Gam_{\lam}{}^k_{0\mu} \dot{x}^\mu 
\\
          && + (\der_\alp X^i) \Gam_{\lam}{}^k_{0\mu} 
                                            \dot{x}^\alp
            - (\der_j X^k) \Gam_{\lam}{}^i_{0\mu} 
                                            \dot{x}^\mu\} 
\,\dot{\der}_k \ten d^\lam\,.\END
\eeq
\ePr

Now let us consider the case when $\Gam$ and $\gam$ are the 
corresponding connections for a spacetime connection $K$.  
The natural bijective correspondences 
\ref{correspondenceKGamgam}
suggest the following theorem.  

\bTh\label{KGamgam}
Let $X$ be an infinitesimal spacetime symmetry. 
The following equivalences hold 
\beq
1)\; L_{X_{(T)}}\,K = 0 
\qquad\Leftrightarrow\qquad
2)\; L_{X_{(1)}}\,\Gam = 0
\qquad\Leftrightarrow\qquad
3)\; L_{X_{(1)}}\,\gam = 0 
\eeq
\eTh

\bPf
The proof of 
$1)\;\Rightarrow\; 2)\;\Rightarrow\; 3)$
follows easily in virtue of the Leibnitz rule and the fact that 
$X_{(1)}$ is a symmetry of the contact maps.
 
The proof of 
$3)\;\Rightarrow\; 2)\;\Rightarrow\; 1)$ 
can be obtained easily by considering the expressions of Lemma
\ref{symexpressions} which are polynomial in the coordinates
$x^i_0$ or $\dot{x}^\mu$.

\vspace{-5mm}\rightline\QED
\ePf

\subsubsection*{Symmetries of phase 2--forms}

Let us consider a spacelike metric $G$ and a phase connection $\Gam$.
Moreover, let $\Ome$ be the corresponding dynamical phase 2--form,
$\gam$ the corresponding dynamical connection and $\cE$ the 
corresponding Euler-Lagrange morphism.  
The natural correspondences 
\ref{correspondenceOmeE}
suggest the following theorem

\bTh\label{OmeGGam}
Let $X$ be an infinitesimal symmetry of spacetime.
The following equivalence holds
\beq
1)\; L_{X_{(1)}}\, \Ome = 0 
\qquad \Leftrightarrow \qquad
2)\; L_{X_{(1)}}\,\Gam = 0\,,\; L_X\, G = 0\,.
\eeq
\eTh

\bPf
The definition \eqref{Omega} yields,
$L_{X_{(1)}}\,\Ome = L_{X_{(1)}}\,\nu[\Gam] \bar\wed \vartheta_1
        + \nu[\Gam] \bar\wed L_{X_{(1)}}\, \vartheta_1
        + \nu[\Gam] \tilde\wed \vartheta_1$.
The proof of $2) \;\Rightarrow\; 1)$ can be seen directly
because of the fact that $X_{(1)}$ is a symmetry of $\vartheta_1$.

For the proof of $1) \;\Rightarrow\; 2)$ we consider
the coordinate expressions 
$L_{X_{(1)}}\,\nu[\Gam] \bar\wed \vartheta_1=
G^0_{ij} \alp^i_{\mu 0} (d^\mu \wed d^j - x^j_0 d^\mu \wed d^0)$
and
$\nu[\Gam] \tilde\wed \vartheta_1=  
   \bet^0_{ij} (        d^i_0 \wed d^j
                - x^j_0 d^i_0 \wed d^0
                - \Gam^i_{\lam 0} d^\lam \wed d^j
                + x^j_0 \Gam^i_{\lam 0} d^\lam \wed d^0)$,
where we have set $\alp^i_{\mu 0}$ to be the coefficient of
$L_{X_{(1)}}\,\Gam$ in Proposition \ref{symexpressions} 
and $\bet^0_{ij}$ the coefficient of $L_X\,G$ in 
Proposition \ref{symexpressionG}. 
It can be easily seen that, if the sum of 
these expressions is zero, then, all $\alp^i_{\mu 0}$ and
all $\bet^0_{ij}$ have to be zero. 

\vspace{-5mm}\rightline\QED      
\ePf 


Moreover, the correspondences \ref{correspondenceOmeE} 
suggest the following
\bTh\label{OmeE}
Let $X$ be an infinitesimal spacetime symmetry. 
The following equivalence holds.
\beq
1)\; L_{X_{(1)}}\,\Ome = 0 \qquad\Leftrightarrow\qquad 
2)\; L_{X_{(2)}}\,\cE = 0 
\eeq
\eTh
                       
\bPf
Expression \ref{EulerLagrange} yields 
$L_{X_{(2)}}\,\cE = L_{X_{(1)}}\,\nab[\gam] \bar\wed \vartheta_1
   + \nab[\gam] \bar\wed L_{X_{(1)}}\,\vartheta_1+ 
     \nab[\gam] \tilde\wed \vartheta_1$.
In analogy to the proof of Theorem \ref{OmeGGam} this yields 
the equivalence between the condition $L_{X_{(2)}}\,\cE = 0$ 
and the two conditions $L_X\,G = 0$, $L_{X_{(1)}}\,\nab[\gam] = 0$.
Clearly, $L_{X_{(1)}}\,\nab[\gam] = 0$ is equivalent to
$L_{X_{(1)}}\,\gam = 0$. Thus, Theorem \ref{OmeGGam} and Theorem
\ref{KGamgam} yield the result.

\vspace{-5mm}\rightline\QED     
\ePf 

Eventually, we add an example of a distinguished 
nonholonomic infinitesimal symmetry of $\Ome$. 
\bPr
The corresponding dynamical connection $\gam$ is a 
nonholonomic (scaled) infinitesimal symmetry of the
cosymplectic structure $(J_1\bE, \Ome, dt)$, 
\ie $L_{\gam}\,\Ome = 0$ and 
$L_{\gam}\,dt = 0$.
\ePr
\bPf
This follows directly from Cartan's formula using 
$i_\gam\,\Ome = 0, i_\gam\,dt = 1$ and the closure of $\Ome$.

\vspace{-5mm}\rightline\QED 
\ePf
We observe that this proposition is essentially the standard 
result for the Reeb vector field of any cosymplectic 
structure applied to our particular case.

But we can even say more about $\gam$.
\bTh
There is exactly one second order connection $\tilde{\gam}$ 
such that  $L_{\tilde{\gam}}\,\Ome = 0$. 
Namely, $\tilde{\gam} = \gam$.
\eTh 

\bPf
Let $\tilde{\gam}$ be a second order connection such that 
$L_{\tilde{\gam}} \Ome  = 0$.
This implies that there exists a local function $f: J_1\bE \to \R$
such that $i_{\tilde{\gam}}\Ome = df$. Using $i_\gam\,\Ome = 0$,
we get $i_{\tilde{\gam} - \gam}\,\Ome = df$.
Let us set $c := \tilde\gam - \gam$; 
$c$ is valued into $\BT ^* \ten \BT ^* \ten V\bE$, and has
coordinate expression $c = c_0{}^i_0 u^0 \ten \der^0_i \,$,
with $c_0{}^i_0 \byd \tilde{\gam}_0{}^i_0 - \gam_0{}^i_0:
J_1\bE \to \R$.
We have to show that a local function $f$ only exists if $c=0$.
Calculating in coordinates using $c_{0i} \byd G^0_{ij} c_0{}^i_0$
one gets the following system of equalities
$\der_0 f = - c_{0i} x^i_0$, $\der_i f = c_{0i}$,
$\der^0_i f = 0$.
This systems implies that $c = 0$. Thus, $\tilde{\gam} = \gam$.

\vspace{-5mm}\rightline\QED
\ePf
\subsection{Noether Symmetries} 
\label{symNoether}

Let us consider a closed dynamical phase 2--form $\Ome$.
The next proposition relates infinitesimal symmetries of
$\Ome$ to conserved quantities.  

\bLm\label{conservedOme}
Let $\Ome$ be a closed dynamical phase 2--form and
let $Y: J_1\bE \to TJ_1\bE$ be an infinitesimal symmetry of $\Ome$.
Then, the 1--form $i_Y\,\Ome$ is closed, and any local 
potential function $f$ of $i_Y\,\Ome$ is a conserved quantity.
\eLm
\bPf
$L_Y\,\Ome = 0$ is locally equivalent to the closure of $i_Y\,\Ome$. 
Hence, there is a local function $f$ such that 
$df = i_Y\,\Ome$. Therefore,
$\gam . f = df( \gam ) = i_Y\,\Ome (\gam) =  
                         i_Y\,i_\gam\,\Ome = 0$.

\vspace{-5mm}\rightline\QED
\ePf 
\subsubsection*{Symmetries of Poincar\'e--Cartan forms}

Let us consider a local Poincar\'e--Cartan form $\The$
associated to $\Ome$.

Clearly, any infinitesimal symmetry $Y: J_1\bE \to TJ_1\bE$ 
of $\The$ is an infinitesimal symmetry of $\Ome$. 
In fact, if $L_Y\,\The = 0$, then 
$0 = dL_Y\,\The = L_Y\,d\The = L_Y\,\Ome$.

Now, we can formulate the following (Noether) theorem which 
relates holonomic infinitesimal symmetries of $\The$ to conserved 
quantities.

\bTh\label{Thepotential}
Let $X$ be an infinitesimal spacetime symmetry which, additionally,
is a holonomic infinitesimal symmetry of $\The$. 
Then, on the domain of $\The$, $i_{X_{(1)}}\,\Ome$ is exact and 
$f \byd - i_X\,\The$ is a potential, hence, a conserved quantity.
\eTh
\bPf
$0 = L_{X_{(1)}}\,\The = (d\,i_{X_{(1)}} + i_{X_{(1)}} d)\, \The
                 = d\,i_{X_{(1)}}\,\The + i_{X_{(1)}}\,\Ome$. 
This is equivalent to the equation  
$i_{X_{(1)}}\,\Ome = - d\,i_{X_{(1)}}\,\The = - d\,i_X\,\The $. 
It follows directly from Lemma \ref{conservedOme} that $- i_X\,\The$ 
is a conserved quantity. 

\vspace{-5mm}\rightline\QED
\ePf

\bRm
In particular, if an observer $o$ is a (scaled) infinitesimal 
symmetry of $\The$, then the Hamiltonian $H[o]$ turns out to
be the associated conserved quantity. 
\eRm

\subsubsection*{Symmetries of Lagrangians}

Let $\cL$ be the Lagrangian corresponding to a Poincar\'e-Cartan 
form $\The$ and let $\cP$ be the corresponding momentum.
The natural correspondence \ref{correspondenceTheL} indicates
the following theorem
\bTh\label{symTheL} 
Let $X$ be an infinitesimal spacetime symmetry.
Then, the following equivalence holds
\beq
1)\; L_{X_{(1)}}\,\The = 0 \qquad\Leftrightarrow\qquad 
2)\; L_{X_{(1)}}\,\cL = 0 
\eeq
\eTh
\bPf
Both directions can be proved in analogy to the proof of the 
equivalence 
$L_{X_{(1)}}\,\Gam = 0\;\Leftrightarrow\;L_{X_{(1)}}\,\gam = 0$.

\vspace{-5mm}\rightline\QED
\ePf

Theorem \ref{symTheL} yields immediately another formulation  
of the (Noether) Theorem \ref{Thepotential}. This version may
be more popular to the physicist. 

\bCr\label{Lpotential}
Let $X$ be an infinitesimal spacetime symmetry 
which, additionally, is a holonomic infinitesimal symmetry of $\cL$.
Then, on the domain of $\The$, a conserved quantity is given 
by 
\beq
f \byd - (X \con \cP + \uX \con \cL)\,.\END
\eeq
\eCr

\myskip 

\subsection{The momentum map in covariant classical mechanics}
\label{momentummap}

Let us suppose a closed dynamical phase 2--form $\Ome$ and a
left action $\hat\Phi: \bG \car J_1\bE \to J_1\bE$
of a group $\bG$ of symmetries of the cosymplectic 
structure $(J_1\bE, \Ome, dt)$. That is, 
$\hat\Phi_g^* \Ome = \Ome$ and $\hat\Phi_g^* dt = dt$. 
Let $\Fg$ be the associated Lie algebra. 
Hence, $L_{\der\hat{\Phi}(\xi)}\,\Ome = 0$ 
and $L_{\der\hat{\Phi}(\xi)}\,dt=0$ for all $\xi \in \Fg$. 

We would like to define a {\em momentum map\/} in our setting by 
analogy with the standard symplectic and cosymplectic literature
\cite{Albert89,dLS93,Mar83,MaRa95} and ref. therein. 

Lemma \ref{conservedOme} shows that any vector field $Y$ of $J_1\bE$
is an infinitesimal symmetry of $\Ome$ if and only if there exists 
a (local) function $f$ such that $i_Y\,\Ome = d\,f$. Clearly, $f$ is 
determined up to an additive constant $c \in \R$. Each $f$ of
this type is a conserved quantity.       

Analogously, we can easily see that the following lemma holds
\bLm\label{consdt}
Let $Y$ be any vector field of $J_1\bE$. Then,  
$Y$ is an infinitesimal symmetry of $dt$ if and only if 
$i_Y\,dt$ is a constant $c \in \bar\BT$.  
\eLm

Hence, by Lemma \ref{conservedOme} and Lemma \ref{consdt}, we can 
locally associate with any infinitesimal symmetry 
$\der\hat\Phi(\xi)$ of $\Ome$ and $dt$ locally a pair 
$( f_\xi ,\tau_\xi)$, where $\tau_\xi$
is the constant $dt(\der\hat\Phi(\xi))$ and $f_\xi$ is a potential 
function of $i_{\der\hat\Phi(\xi)}\,\Ome$.  
In the following we denote by $\Co(J_1\bE)$ the vector space of
conserved quantities.

\bDf
A (local) map $J$ 
\beq
J: \Fg \to \Co(J_1\bE) \car \bar\BT :
\,\xi \mto (J_\xi, \tau_\xi)\,,
\eeq
where $J_\xi$ is a potential of $i_{\der\hat\Phi(\xi)}\,\Ome$
and $\tau_\xi \byd i_{\der\hat\Phi(\xi)}\,dt$ for all $\xi \in \Fg$,  
is said to be a {\em momentum map} for the action $\hat{\Phi}$.
\eDf

\bRm
In general, a momentum map $J$ is defined locally. 
But if we assume suitable hypotheses on spacetime or on the 
Lie algebra $\Fg$, then we can find a global momentum map. 
Of course, a global $J$ always exists if
$H^1(\bE) = \{0\}$. A detailed list of other hypotheses under which
$J$ is globally defined is given in \cite{MaRa95}; 
they are the same as in our case.\END
\eRm

On the other hand, given a time scale $\tau$, it is possible to 
associate with any function $f$
of $J_1\bE$ a distinguished vector field of $J_1\bE$, 
namely, the $\tau$-Hamiltonian lift of $f$. 
The following theorem holds 

\bTh\label{genlift}
Let $J$ be a momentum map for the action $\hat\Phi$ and
let $H_\tau[J_\xi]$ be the $\tau$-Hamiltonian lift of $J_\xi$ 
with respect to an arbitrary time scale $\tau$.
 
Then, the following equivalence holds
\beq
\der\hat\Phi(\xi) =  H_\tau[J_\xi]\,\Leftrightarrow\, 
\tau \byd \tau_\xi = i_{\der\hat\Phi(\xi)}\,dt.
\eeq 
\eTh 
\bPf
By recalling that $\gam. J_\xi = 0$ we obtain that
$i_{H_{\tau}[J_\xi]}\,\Ome = dJ_\xi - \langle \gam. J_\xi, dt\rangle
 = dJ_\xi$.  
Hence, by observing that two vector fields $X,\,Y: J_1\bE \to TJ_1\bE$
are equal if and only if $i_X\,dt = i_Y\,dt$ and 
$i_X\,\Ome = i_Y\,\Ome$, we obtain the result.  

\vspace{-5mm}\rightline\QED
\ePf

This theorem shows, why we have included the time scale $\tau_\xi$
in the definition of momentum map. Again, we stress the fact, that 
the function $J_\xi$ is only determined up to a gauge, whereas the
time scale $\tau_\xi$ is determined uniquely by $\der\hat\Phi(\xi)$.

It is obvious that we want to know if the map that associates
to a pair $(J_\xi, \tau_\xi)$ its $\tau$-Hamiltonian lift
${\it H}(J_\xi, \tau_\xi) \byd H_{\tau_\xi}[J_\xi]$ is a homomorphism 
of Lie algebras. Therefore, we define the following bracket for pairs
in $\Co(J_1\bE) \car \bar\BT$.
\bDf 
The bracket 
$\{(f,\tau),\,(g,\sig)\} \byd (0,\,\{f,\,g\})$
is said to be the {\em Poisson bracket of pairs}.
\eDf 

Then we can prove the following theorem
\bTh
The map ${\it H}$ is a homomorphism of Lie algebras between pairs
$(f, \tau)\,\in \Co(J_1\bE) \car \bar\BT$, with respect to the
Poisson bracket of pairs, and vector fields 
$H_\tau[f]$ of $J_1\bE$, with respect to the standard Lie bracket.
\eTh
\bPf
The equalities
$i_{[H_\tau[f], H_\sig[g]]}\,dt = 
 [L_{H_\tau[f]}, i_{H_\sig[g]}]\,dt = 
  L_{H_\tau[f]}\,\sig - i_{H_\sig[g]}\,d\tau = 0$
and
$i_{[H_\tau[f], H_\sig[g]]}\,\Ome = 
 [L_{H_\tau[f]}, i_{H_\sig[g]}]\,\Ome = 
 d\,i_{H_\tau[f]}\,i_{H_\sig[g]}\,\Ome
+ i_{H_\tau[f]}\,L_{H_\sig[g]}\,\Ome
- i_{H_\sig[g]}\,L_{H_\tau[f]}\,\Ome = d\{f, g\}$.
yield the result.

\vspace{-5mm}\rightline\QED
\ePf

Let us recall that the $\tau$-Hamiltonian lift of a function 
$f: J_1\bE \to \R$
is projectable on a vector field $X$ of $\bE$ if and only if
$f$ is a special function and the second fibre derivative of $f$ 
(with respect to the velocities) is equal to the time scale $\tau$.
Thus, Theorem \ref{genlift} yields the following theorem that 
relates the function $J_\xi$ to the time scale $\tau_\xi$.
 
\bCr
Let $\hat\Phi$ be projectable on a left action 
$\Phi: \bG \car \bE \to \bE$. Then, any component $J_\xi$ of a 
momentum map for $\hat\Phi$ is a quantisable function and the second 
fibre derivative of $J_\xi$ is equal to the time scale 
$\tau_\xi = dt(\hat\Phi(\xi))$. 
\eCr
Thus, in this case, each function $J_\xi$ encodes all information of 
the pair $(J_\xi, \tau_\xi)$. Hence, we call the map
$J: \Fg \to \Co(J_1\bE): \xi \to J(\xi) \byd J_\xi$
momentum map, denoted by the same symbol $J$.
   
Now let us consider a Poincar\'e--Cartan form $\The$. 
Furthermore, let us suppose that $\hat\Phi$ is holonomic, \ie
$\hat\Phi = \Phi_{(1)}$ where $\Phi$ is a left action of $\bG$ on 
$\bE$ and, additionally, we suppose that $\Phi$ preserves
$\The$. Then the following holds

\bTh
There exists a momentum map on the domain of $\The$. 
Namely, the map
\bEq
J_\xi = \der\Phi(\xi) \con \cP + \underline{\der\Phi}(\xi) 
\con L\,.
\eEq

Let $(e_p)$ be a basis of $\Fg$, and $\xi = \xi^pe_p$. 
Then, the coordinate expression is
\beq
J_\xi = \xi^p\big((\der_p\phi^i -x^i_0\der_p\phi^0) \, \der^0_iL + 
\der_p\phi^0\, L\big)\,,
\eeq

Given an observer $o$, the momentum map can be expressed in terms
of the observed Hamiltonian $\cH[o]$ and the observed momentum
$\cP[o]$ by
\bEq
J_\xi = \der\Phi(\xi) \con \cP[o] + \underline{\der\Phi}(\xi) 
\con \cH[o]\,.
\eEq
\eTh

\bPf
The first expression follows simply from the contact splitting of 
$\The$ and Theorem \ref{Thepotential}.
The observer dependent expression of $J$ follows simply from the 
splitting of $\The$ through the observer.

The coordinate expression 
$\der\phi(\xi) = \xi^p\der_p\phi^0\der_0 + 
\xi^p\der_p\phi^i\der_i$ with respect to a basis $(e_p)$ yields the 
second expression.

\vspace{-5mm}\rightline\QED
\ePf

\bRm\label{momentum}
There is a connection between the momentum of a Lagrangian and the 
momentum map. In fact, let $\bG$ be a group of vertical holonomic
symmetries of $\The$, \ie $i_{\der\Phi(\xi)}\,dt = 0$. 
Then we have the expression
\beq
J_\xi = \der\phi(\xi) \con \cP \equiv 
\cP (\der\phi(\xi)) \,,
\eeq
so the momentum map coincides with the momentum of the Lagrangian.\END
\eRm

\subsection{General model of {\em CCG\/}}

The general model of {\em CCG\/} is constituted by a spacetime $t:\bE \bT$,
a spacelike metric $G$, a spacetime connection $K\Nat$
(gravitational field) and a scaled 2--form 
$f: \bE \to (\BL^{1/2} \ten \BM^{1/2}) \ten \Lam^2 T^*\bE$ 
of spacetime (electromagnetic field).

The gravitational connection $K\Nat$ yields the gravitational 
objects $\Gam\Nat$, $\gam\Nat$ and $\Ome\Nat$ through the
correspondences discussed in the above section:
$K \leftrightarrow \Gam[K] \leftrightarrow \gam[\Gam]$ and 
$\Gam \leftrightarrow \Ome[G, \Gam]$.
 
On the other hand, given a charge $q \in \BQ$, it is convenient to 
introduce the normalised form 
$F \byd \frac{q}{m}f: \bE \to \Lam^2 T^*\bE$.
This form can be regarded naturally as a dynamical phase 2--form.

The first field equations for the gravitational and 
the electromagnetic field are assumed to be 
$d \Ome\Nat = 0$ and $dF = 0$.

There is a natural way to incorporate the electromagnetic
field in the gravitational objects, by means of {\em total objects}
of the type $K = K\Nat + K^e$, $\;\Gam = \Gam\Nat + \Gam^e$,
$\;\gam = \gam\Nat + \gam^e$ and $\;\Ome = \Ome\Nat + \Ome^e$.
In fact, we can start by considering a natural ``minimal coupling''
\cite{Kun84,DuKu84}
\bEq
\Ome \byd \Ome\Nat + \frac12 F \,,
\eEq
where the factor $\frac12$ has been chosen just in order to obtain
standard normalisation.
Then the other total objects are obtained by means of the 
correspondences  
$K \leftrightarrow \Gam[K] \leftrightarrow \gam[\Gam]$ and 
$\Gam \leftrightarrow \Ome[G, \Gam]$.
In particular, $-\gam^e$ turns out to be the Lorentz force.
The coefficients of $K$ turn out to be
$K{_h}{^i}{_k}=K\Nat{_h}{^i}{_k}$,
$K{_0}{^i}{_k}=K\Nat{_0}{^i}{_k}+\frac{q}{2m} F{^i}{_k}$ and
$K{_0}{^i}{_0}=K\Nat{_0}{^i}{_0}+\frac{q}{m} F{^i}{_0}$. 

This yields that $dt\wed\Ome^n = dt\wed(\Ome\Nat)^n : 
J_1\bE \to \BT \ten \Lam^n T^*J_1\bE$ is a (scaled) volume form
of $J_1\bE$ and that $d\Ome = 0$. 
Therefore, the phase space $J_1\bE$ together with the total 
phase 2--form $\Ome$ and the time form $dt$ turns out to be a 
(scaled) cosymplectic manifold \cite{Albert89,CdLL92,dLS93}. 
The total dynamical connection $\gam$ turns out to be the 
(scaled) Reeb vector field for this cosymplectic structure
since $i_\gam\,\Ome = 0$ and $i_\gam\,dt = 1$.  

\subsection{Examples}
Now, we apply the machinery developed in the above 
subsection to analyse three groups of symmetries acting in simple 
cases. 

\bEx
We suppose the spacetime $\bE$ to be an affine space with affine 
projection $t$. In this case $V\bE \simeq \bE\car\bS$, where
$\bS \byd \ker Dt$. So, we assume an Euclidean scaled metric $g$
on $\bS$.

Let us consider the vertical action
\beq
\bS \car \bE \to \bE : (v, e_0) \mto (e_0+v)\,;
\eeq

Let $K\Nat$ be the natural flat connection on $\bE$ and $F = 0$.
Then, any Poincar\'e--Cartan form exists globally and 
$\bS_0$ is a group of symmetries of a $\The$. The 
momentum map $J$ is just the standard linear momentum.

In fact, $\The$ is invariant with 
respect to spacelike translations. 
Of course, the Lie algebra of $\bS_0$ 
is $\bS_0$ and we have the momentum map
\beq
J : \bS_0 \to C^{\infty}(J_1\bE , \R) 
  : v \mto J(v) \equiv p_{L}(v)\,.
\eeq
We have the coordinate expression  
$p_{L_0}(v) = v^i \, G_{ij}x^j_0$
(see remark \ref{momentum}).\END
\eEx

\bEx
Assume the same spacetime and fields as in the above example, and
assume additionally that $\bE \simeq \bT \car \bP$, \ie, 
assume a complete observer $o$. Then, we can consider the natural 
action
\beq
\olin{\BT} \car (\bT \car \bP) \to \bT \car \bP :
(v, (\tau, \boldsymbol{p})) \mto (v+\tau,\boldsymbol{p}) \,.
\eeq

It turns out that $\olin{\BT}$ is a group of symmetries 
of $\The$, \ie $o$ is a (scaled) infinitesimal symmetry
of $\The$, and the momentum map $J$ is just the (observed) kinetic 
energy $\cH[o]$.

In fact, $\The$ is as in the above example, hence it 
is invariant with respect to time translations because the metric 
does not depend on time. Of course, the Lie algebra of 
$\olin{\BT}$ is $\olin{\BT}$ and we have the momentum map
\beq
J : \olin{\BT} \to \Co(J_1\bE) :  \xi \mto 
J(\xi) \equiv \xi \con (o \con \The)\,.
\eeq
Obviously, $J = \cH[o]$.\END
\eEx

\bEx
Now, we suppose our spacetime to be $\bT \car SO(g)$,
where $g$ is the metric of the above spacetime. The manifold 
$\bT \car SO(g)$ is interpreted as the configuration space for the
relative configurations of a rigid body with respect to the center of
mass (see \cite{MaRa95,MTV99} for a more detailed account).

We assume the {\em inertia tensor\/} $I$ as the scaled vertical 
metric. Consider the action
\beq
SO(g) \car (\BT \car SO(g)) \to \BT \car SO(g) :
(A, (\tau, B)) \mto (\tau,AB) \,.
\eeq

Let $K\Nat$ be the natural flat connection on $\bT \car SO(g)$ and
$F=0$. Then, $SO(g)$ is a group of symmetries of $\The$ and a
momentum map $J$ is just the angular momentum.

In fact, as in the previous examples, $\The$ reduces to the kinetic 
energy of particles with respect to the center of mass. 
This is obviously invariant with respect orthogonal transformations 
\cite{MTV99}. 
We have the momentum map
\beq
J : so(g_a) \to \Co(\bT \car \BT\ten TSO(g)) :  \ome \mto 
J_\ome \equiv \ome^*\con \cP\,,
\eeq
where, by definition, $\cP = V_{\bE}L$, with the coordinate 
expression $\cP = I_{ij}x^j_0\check{d}^i$. A simple computation 
shows that

-- $\ome^* : SO(g) \to TSO(g) : r \mto \ome(r)$;

-- $\ome^*\con \cP(v) = I(\ome(r),v) = \ome (r \car v)$.

The Lie algebra of $SO(g_a)$ is $so(g_a)$, but the Hodge star 
isomorphism yields a natural Lie algebra isomorphism $so(g_a) 
\simeq \BL^{-1}\otimes\bS_a$. The isomorphism carries the Lie 
bracket of $so(g_a)$ into the cross product. 
In this way, if $\ome \in so(g_a)$ and 
$\Bar{\ome}\in \BL^{-1}\otimes\bS_a$ is the corresponding element, 
we can equivalently write
\beq
J : \BL^{-1}\otimes\bS_a \to \Co(\bT \car \BT\ten TSO(g)) :  
\Bar{\ome} \mto J_{\Bar{\ome}} \equiv I(r \car v,\ome)\,,
\eeq
where $v \in \BT^* \otimes T\bR_a \equiv J_1(\bT \car\bR_a)$. This 
proves the last part of the statement.\END
\eEx

\section{Conclusions}

In this paper we used natural bijective correspondences 
between distinguished objects in the covariant classical Galilei 
theory as an indicator for relations between symmetries of these 
objects. In particular, it turned out that the holonomic 
infinitesimal symmetries of the objects spacelike metric, 
gravitational field and electromagnetic field are equivalent
to the holonomic infinitesimal symmetries of induced
cosymplectic structure. Moreover, these symmetries are equivalent to 
the holonomic infinitesimal symmetries of the associated 
Euler-Lagrange morphism.
Analogously, we saw that the holonomic infinitesimal symmetries 
of the horizontal potential 1-forms of the cosymplectic form 
are equivalent to the holonomic infinitesimal symmetries of the
associated Lagrangians. 
These results allowed us to give two equivalent versions of
the Noether theorem, namely, a `Poincar\'e--Cartan' version and
a `Lagrangian' version. 
Thus, we saw that the holonomic infinitesimal symmetries of 
the cosymplectic structure yield the holonomic infinitesimal 
symmetries of all physical objects and of the dynamics.

Then, we introduced a covariant momentum map in the model, associated
with cosymplectic symmetries. Here, the covariance with respect to
units, lead us to introduce a momentum map whose components are pairs, namely, a constant time scale and a conserved quantity. A momentum map was determined only up to an additive constant to the conserved quantities. We saw that any pair determines uniquely a covariant lift ($\tau$-Hamiltonian lift) of functions. We introduced a Poisson bracket for these pairs, such that the covariant lift turned out to be a morphism of Lie algebras. Then, we showed that for any action of cosymplectic symmetries which projects on an action on spacetime, the components of the momentum map turned out to be `quantisable functions' and their second fibre derivative determined the constant time scale. This feature is new with respect to standard symplectic momentum map formalism, and is mainly due to covariance requirements. Its importance is fundamental in quantum mechanics. We saw that, if the group of symmetries , additionally, preserves a given Poincar\'e--Cartan form, then the momentum map is uniquely determined by the Poincar\'e--Cartan form.
   
Finally, we provided simple examples illustrating our results at work.

\myskip

These results are promising in view of a next research about 
symmetries of quantised systems.

\end{document}